\documentclass[aps,prl,reprint,superscriptaddress]{revtex4-1}
\usepackage{graphicx}
\usepackage[dvipsnames]{xcolor}
\usepackage{multirow}
\usepackage{amsmath}
\usepackage[normalem]{ulem}

\newif\ifRPSHighlitedChanges
\def\ifRPSHighlitedChanges{\iftrue}
\ifRPSHighlitedChanges
  
  \def\SRPS#1{{\color{Purple}\sout{#1}}}
  \def\CRPS#1{{\color{Purple}\textbf{(Comment: #1)}}}
  
\else
  
  \def\SRPS#1{\relax}
  
  \def\CRPS#1{\relax}
\fi

\newif\ifCOHighlitedChanges
\def\ifCOHighlitedChanges{\iftrue}
\ifCOHighlitedChanges
  
  \def\SCO#1{{\color{Orange}\sout{#1}}}
  \def\CCO#1{{\color{Orange}\textbf{(Comment: #1)}}}
  
\else
  
  \def\SCO#1{\relax}
  
  \def\CCO#1{\relax}
\fi

\begin{document}

\title{Yielding and bifurcated aging in nanofibrillar networks}

\author{Ryan Poling-Skutvik}
\affiliation{Department of Chemical and Biomolecular Engineering, University of Pennsylvania, Philadelphia, Pennsylvania 19104}

\author{Eoin McEvoy}
\affiliation{Center for Engineering Mechanobiology, University of Pennsylvania, Philadelphia, Pennsylvania, 19104}
\affiliation{Department of Materials Science and Engineering, University of Pennsylvania, Philadelphia, Pennsylvania 19104}

\author{Vivek Shenoy}
\affiliation{Center for Engineering Mechanobiology, University of Pennsylvania, Philadelphia, Pennsylvania, 19104}
\affiliation{Department of Materials Science and Engineering, University of Pennsylvania, Philadelphia, Pennsylvania 19104}

\author{Chinedum O. Osuji}
\email{cosuji@seas.upenn.edu}
\affiliation{Department of Chemical and Biomolecular Engineering, University of Pennsylvania, Philadelphia, Pennsylvania 19104}

\date{\today}

\begin{abstract}
The yielding of disordered materials is a complex transition involving significant changes of the material's microstructure and dynamics. After yielding, many soft materials recover their quiescent properties over time as they age. There remains, however, a lack of understanding of the nature of this recovery. Here, we elucidate the mechanisms by which fibrillar networks restore their ability to support stress after yielding. Crucially, we observe that the aging response bifurcates around a critical stress $\sigma_\mathrm{c}$, which is equivalent to the material yield stress. After an initial yielding event, fibrillar networks subsequently yield faster and at lower magnitudes of stress. For stresses $\sigma<\sigma_\mathrm{c}$, the time to yielding increases with waiting time $t_\mathrm{w}$ and diverges once the network has restored sufficient entanglement density to support the stress. When $\sigma > \sigma_\mathrm{c}$, the yield time instead plateaus at a finite value because the developed network density is insufficient to support the applied stress. We quantitatively relate the yielding and aging behavior of the network to the competition between stress-induced disentanglement and dynamic fluctuations of the fibrils rebuilding the network. The bifurcation in the material response around $\sigma_c$ provides a new possibility to more rigorously localize the yield stress in disordered materials with time-dependent behavior. 
\end{abstract}

\maketitle

Fibrillar networks form the mechanical backbone of many biological materials. Collagen \cite{Jansen2018} and cellulose \cite{Cosgrove2018}, for example, provide the elasticity and rigidity of the extracellular matrix and tissues in animals \cite{Hiltner1985} and the cell wall in plants \cite{Ding2012,Bidhendi2019}, respectively. To accommodate cellular growth and expansion, these fibrillar networks must repeatedly break and restructure \cite{Cosgrove2000,Cosgrove2016,Bidhendi2016}, resulting in complex changes to the material structure and mechanical properties over time. Alternatively, fibrillar networks are often processed under applied stresses, for example, as scaffolds for tissue engineering and regeneration \cite{Miranda-Nieves2017}. Under these processing conditions, external stresses distort the local structure of the fibrillar network and force the material to flow.

How materials transition from an elastic solid to flowing as a viscous liquid (i.e. the yield transition) has been investigated for over a century but the physical underpinnings continue to be debated. Many investigations relate the yield transition to a specific yield stress $\sigma_\mathrm{y}$ or yield strain ${\gamma}_\mathrm{y}$ that causes the material to deviate from a linear elasticity by displaying unrecoverable strain. Although yielding is phenomenologically simple to comprehend, the quantities $\sigma_\mathrm{y}$ and ${\gamma}_\mathrm{y}$ are often defined empirically and depend on the measurement technique \cite{Dinkgreve2016,Donley2019}. Recent work indicates that plastic rearrangements around soft structural defects control yielding in disordered solids \cite{Cubuk2017,Bouzid2018} whereas the dynamic alignment of stress elements controls the degree to which materials can elastically recover strain across the yielding transition \cite{Lee2019}. 
Although the structural and dynamic precursors of the yielding transition have been investigated \cite{Vasisht2018,Rogers2018,Cipelletti2020}, the opposite transition from a flowing fluid back into an elastic solid remains poorly understood.

In this Letter, we show that fibrillar suspensions form physical networks that abruptly yield when subjected to external stresses larger than the yield stress $\sigma_\mathrm{y}$. Because yielding results in a local destruction of the network, subsequent applications of stress result in yielding events that depend strongly on the newly applied stress $\sigma$ and the waiting time $t_\mathrm{w}$ since the initial yielding event. For a given $t_\mathrm{w}$, the time $t_\mathrm{y}$ at which the fibrillar networks yields decreases with increasing $\sigma$. For a constant $\sigma$, $t_\mathrm{y}$  increases with increasing $t_\mathrm{w}$, but the nature of this increase bifurcates around a critical stress $\sigma_\mathrm{c}$. When $\sigma < \sigma_\mathrm{c}$, the yield time $t_\mathrm{y}$ diverges at a finite $t_\mathrm{w}$, indicating that the network structure recovers sufficiently to support the externally applied stress. By contrast, when $\sigma > \sigma_\mathrm{c}$, the yield time plateaus with increasing $t_\mathrm{w}$ because the applied stress exceeds the strength of the equilibrium network structure. We relate the kinetics of the restructuring and yielding phenomena to the segmental dynamics of the fibrils that restores the physically entangled network and stress-activated disentanglement. We observe that $\sigma_\mathrm{c} \approx \sigma_\mathrm{y}$, which suggests that the bifurcation of the aging response may provide a more robust identification of the yield stress in this class of materials. From these findings, we develop a comprehensive physical picture describing the behavior of soft, amorphous materials across the yielding transition as a continuous competition between the destruction and restoration of local microstructure. 

A stock suspension of TEMPO-modified cellulose nanofibrils (CNF) was acquired from the Process Development Center at the University of Maine at a concentration of \hbox{$1.1$ wt.\%} and a surface charge concentration of 1.5 mM per gram of dry CNF. The stock solution was diluted by mixing with deionized water and bath sonicating for 10 min or concentrated using a rotary evaporator. All samples were visually uniform and transparent, indicating that the CNF remained homogeneously suspended. Volume fractions $\phi$ are estimated from weight fractions assuming the specific gravity of cellulose nanofibrils $s = 1.5$. Oscillatory and steady shear rheological measurements were performed on a strain-controlled ARES G2 with a 25 mm cone geometry with an angle of \hbox{0.1 rad}. Creep measurements were performed on a stress-controlled Anton-Paar MCR301 rheometer with a 50 mm parallel plate geometry and a gap size of 1 mm. Sample edges were coated with a thin layer of light mineral oil to prevent evaporation. During the creep experiments, samples are yielded by either applying a constant stress $\approx 2\sigma_\mathrm{c}$ or under a constant shear rate $\dot{\gamma} = 1$ rad s$^{-1}$ until the sample reaches a strain between 500 and 1000\%. After this yielding protocol, the sample is allowed to rest under zero external stress for a waiting time $t_\mathrm{w}$ after which a stress $\sigma$ is imposed and the sample compliance is measured. 

\begin{figure}[bt!]
\includegraphics[width = 3 in]{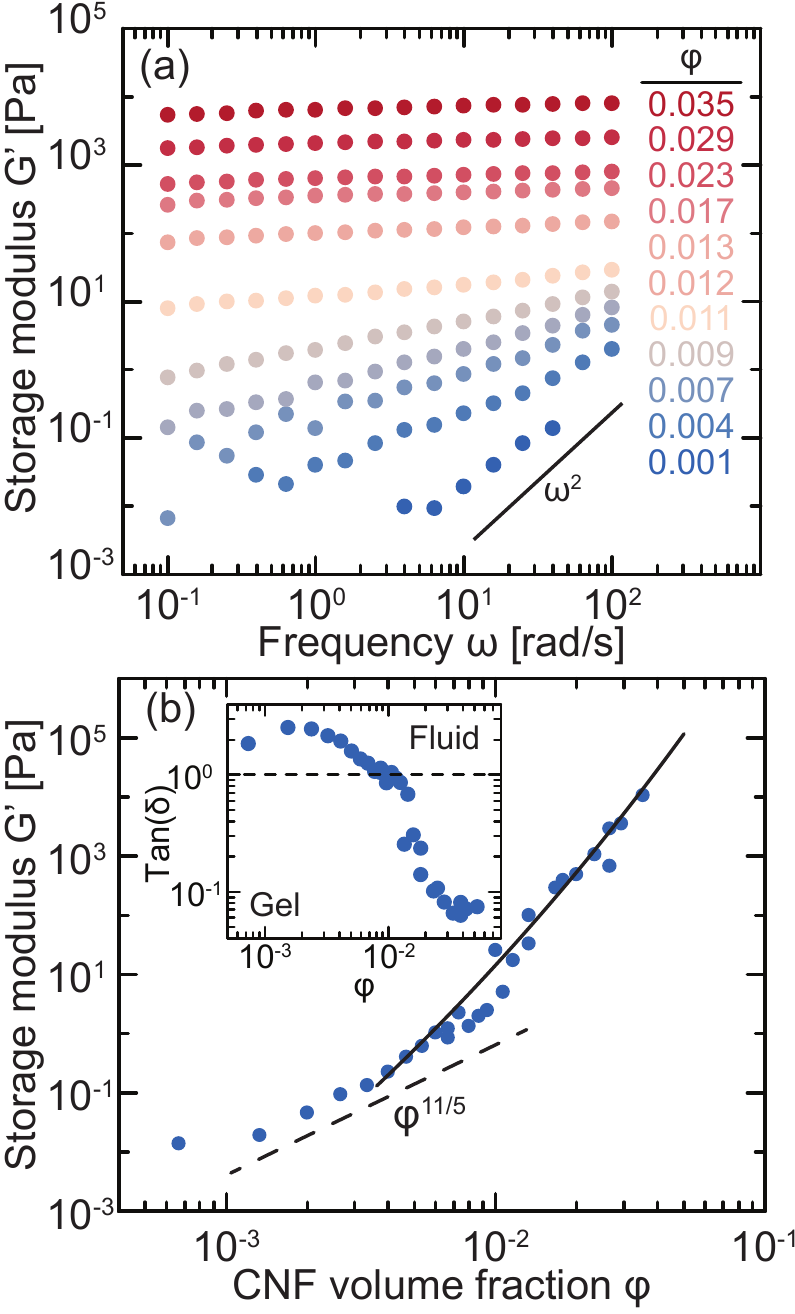}
\caption{\label{fig:freq} Storage modulus $G'$ for fibrillar suspensions as a function of (a) frequency at various concentrations and (b) concentration at $\omega = 10$ rad s$^{-1}$. Dashed curve represents predicted scaling for suspensions of semiflexible fibers \cite{MacKintosh1995} and solid curve indicates predictions for fibrillar networks \cite{Hill2008}. \textit{Inset}: Ratio of moduli $\tan(\delta) \equiv G''/G'$ at $\omega = 10$ rad s$^{-1}$. }
\end{figure}

We characterize how CNF form networks by measuring the rheological properties of the fibrillar suspensions (Fig.\ \ref{fig:freq}). The rheology of the suspensions originates due to physical interactions akin to entanglements, and to weak association involving Van der Waals interactions and hydrogen bonding \cite{Nechyporchuk2016,Xu2018}. At low volume fractions $\phi \lesssim 0.01$, the CNF form viscoelastic fluids in which the storage modulus scales as $G' \sim \phi^{11/5}$ in good agreement with predictions for suspensions of semiflexible fibers that account for stretching entropy and fibrillar bending rigidity \cite{MacKintosh1995}. At higher concentrations, the suspensions form viscoelastic gels (i.e. $\tan(\delta) < 1$) in which $G'$ increases more rapidly with $\phi$. The rapid increase in $G'$ with $\phi$ is consistent with previous investigations \cite{Paakko2007,Quennouz2016} and with a recent model in which elastic energy is stored in deflections of simply supported beams \cite{Hill2008}.
Additionally, the fact that the modulus is nearly independent of oscillation frequency (Fig.\ \ref{fig:freq}) at high $\phi$ indicates that terminal relaxations of the network are suppressed. 

\begin{figure}[tb!]
\includegraphics[width = 3 in]{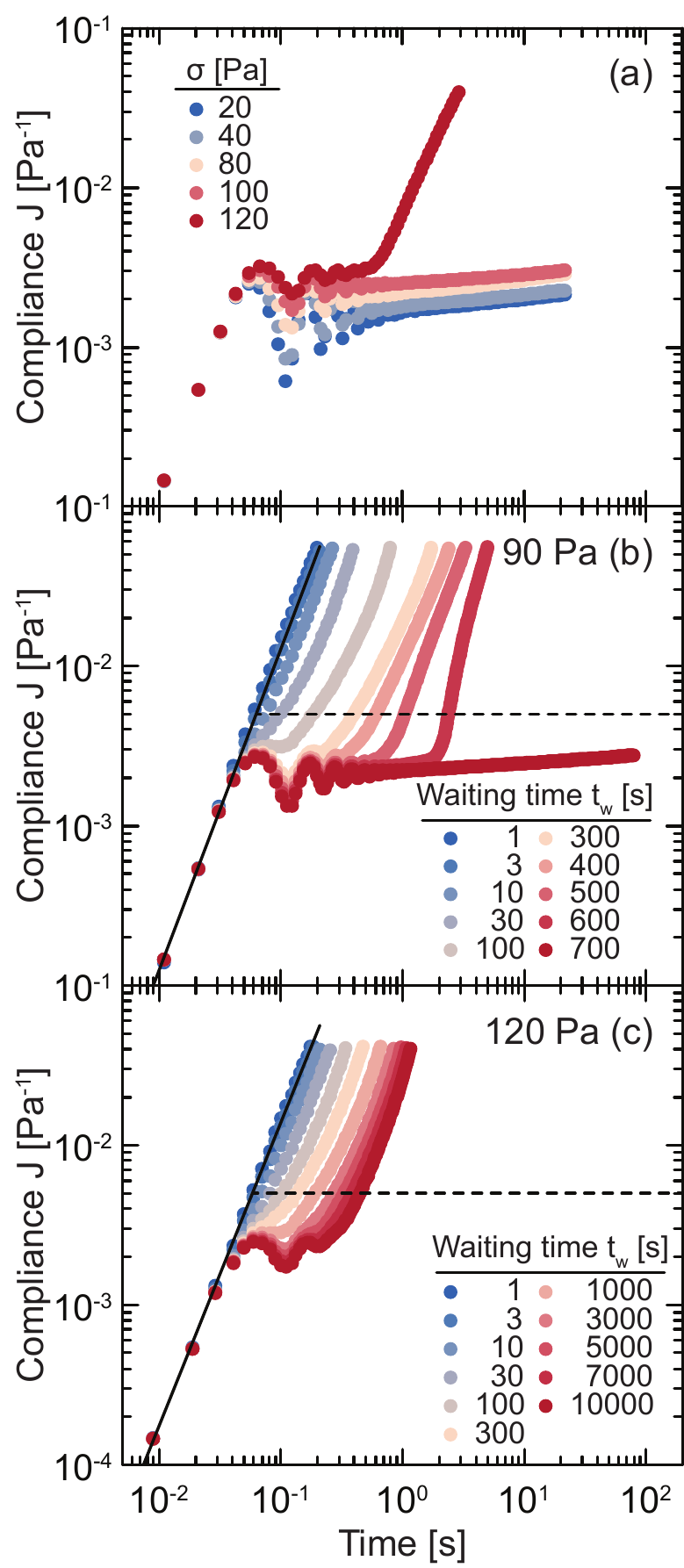}
\caption{\label{fig:yield} (a) Compliance $J = \gamma/\sigma$ as a function of time for a fully equilibrated $\phi = 0.023$ fibrillar network under  different stresses. Compliance $J$ of the network after an initial yielding event with applied stress of (b) 90 Pa and (c) 120 Pa at various waiting times $t_\mathrm{w}$. Solid lines indicate inertial response of instrument. Dashed lines indicate yielding thresholds.}
\end{figure}

To characterize the response of the fibrillar networks under stress, we measure the compliance of fully equilibrated CNF networks as a function of time (Fig.\ \ref{fig:yield}a). When subjected to low stresses, the CNF networks exhibit finite compliances after creep ringing, indicating that the gels behave like elastic solids. Under large external stresses, however, the compliance initially plateaus on short time scales before increasing abruptly as the network begins to flow. The stress at which this behavior occurs is nominally the yield stress $\sigma_\mathrm{y}$. If allowed to come to rest after yielding and then subjected to a second stress, the CNF network will yield rapidly even for stresses $\sigma \ll \sigma_\mathrm{y}$ (Fig.\ \ref{fig:yield}). As the waiting time $t_\mathrm{w}$ increases between a yielding event and subsequent stress application, the CNF network initially supports the stress on short time scales to delay yielding at a characteristic yield time $t_\mathrm{y}$. The delay in yielding depends strongly on the magnitude of the stress applied after initial yielding. For small stresses (e.g. $\sigma = 90$ Pa for $\phi = 0.023$), $t_\mathrm{y}$ increases with increasing $t_\mathrm{w}$ until the sample no longer yields and instead exhibits a finite compliance at long times. Physically, this finite compliance indicates that the CNF network has undergone kinetic restructuring to rebuild a dense enough entanglement network to support the applied stress without large-scale flow. By contrast, the CNF network always yields under high applied stresses (e.g. $\sigma = 120$ Pa for $\phi = 0.023$) even as $t_\mathrm{y}$ moderately increases over many decades of $t_\mathrm{w}$. 

\begin{figure}[tb!]
\includegraphics[width = 3 in]{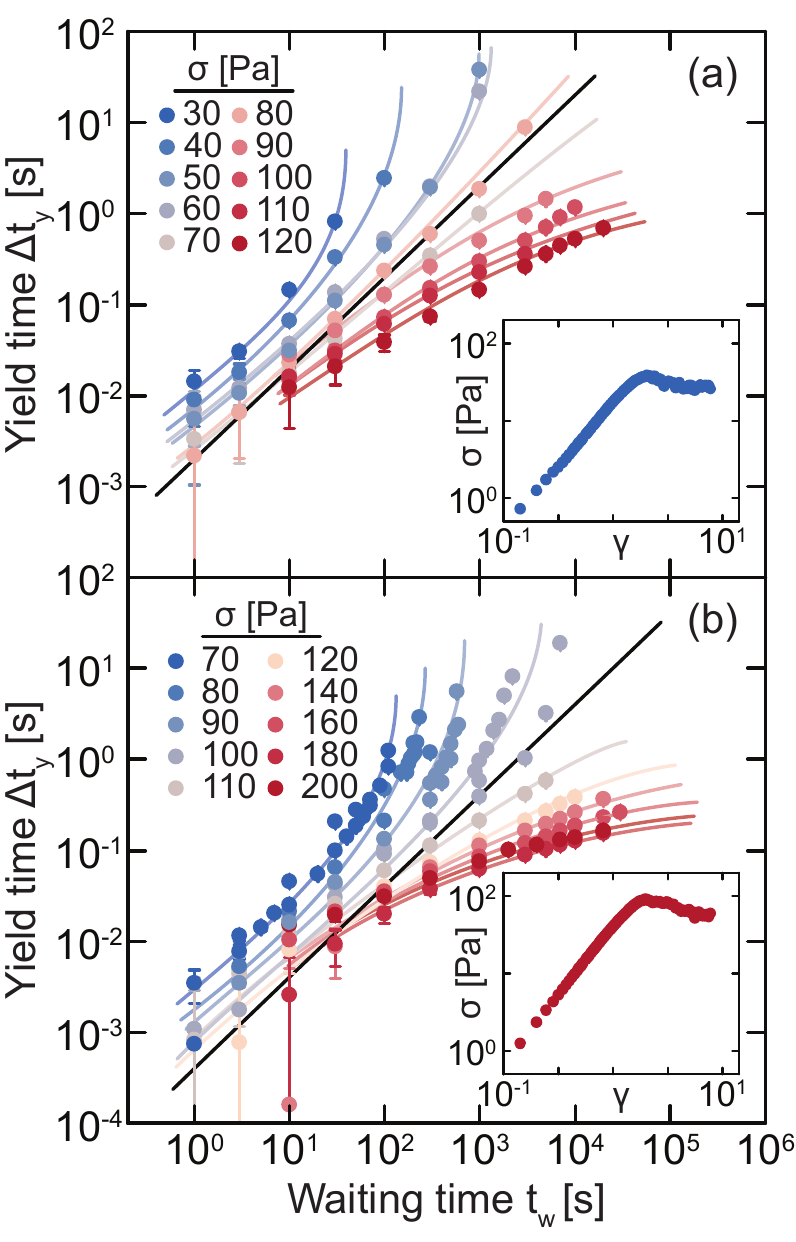}
\caption{\label{fig:aging} Yield time $\Delta t_\mathrm{y}$ as a function of waiting time $t_\mathrm{w}$ under different applied stresses for fibrillar suspensions with (a) $\phi = 0.018$ and (b) $\phi =0.023$ . Colored curves are guides to the eye. Black curves illustrate critical bifurcation behavior. \textit{Insets}: Stress $\sigma$ as a function of strain $\gamma$ measured at a constant shear rate of $\dot{\gamma} = 10^{-3}$ rad s$^{-1}$. }
\end{figure}

We quantify the temporal shift in the compliance curves by defining the yield time $\Delta t_\mathrm{y}$ as the difference between the instrument's inertial response and the time when $J = 2J^*$ (dashed lines in Fig.\ \ref{fig:yield}), where $J^*$ is the height of the first creep ringing peak. This definition unambiguously distinguishes between the yielding flow of the fibrillar network and the creep ringing of the instrument, which is especially important when considering yielding at low $t_\mathrm{w}$. For all samples and all applied stresses, $\Delta t_\mathrm{y}$ increases with $t_\mathrm{w}$ but depends strongly on volume fraction and applied stress. The increase in $\Delta t_\mathrm{y}$ with increasing $t_\mathrm{w}$ indicates that the material dynamically changes under quiescent conditions, but because $t_\mathrm{w} \gg \Delta t_\mathrm{y}$, the CNF networks are at quasi-steady state during each creep experiment. There is a distinct bifurcation in the behavior of $\Delta t_\mathrm{y}$ across the critical stress $\sigma_\mathrm{c}$. For $\sigma < \sigma_\mathrm{c}$, $\Delta t_\mathrm{y}$ increases concavely with $t_\mathrm{w}$ until it diverges, at which point the sample supports the applied stress without bulk flow. As $\sigma$ increases, the divergence of the yield time curve occurs at larger $t_\mathrm{w}$, indicating that it takes longer for the CNF network to dynamically restructure enough to accommodate the larger stress. By contrast, $\Delta t_\mathrm{y}$ increases convexly with $t_\mathrm{w}$ when $\sigma > \sigma_\mathrm{c}$ to reach a plateau. The magnitude of this plateau decreases with increasing $\sigma$, indicating that higher stresses yield the networks more quickly. Thus, the critical stress $\sigma_\mathrm{c}$ separates the solid-like and liquid-like responses of the network and closely agrees with the maximum stress of a fully equilibrated sample under a transient flow sweep (insets to Fig.\ \ref{fig:aging}). This maximum stress often serves as an empirical definition of $\sigma_\mathrm{y}$. By demarcating the boundary between elastic deformation and viscous flow, $\sigma_\mathrm{c}$ serves as a more robust metric to quantify the yield transition so that we can define $\sigma_\mathrm{y} \equiv \sigma_\mathrm{c}$.

\begin{figure}[tb!]
\includegraphics[width = 3.25 in]{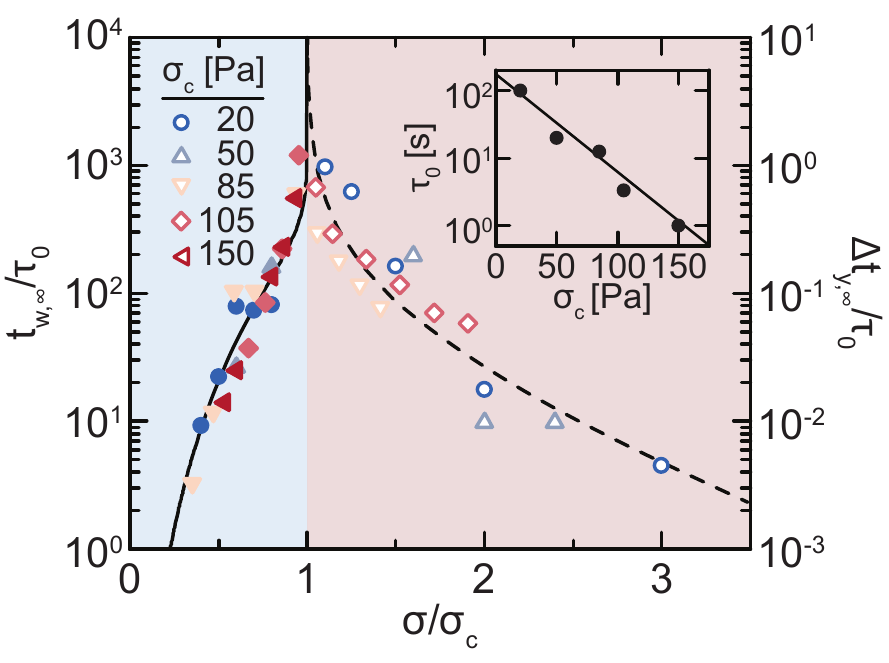}
\caption{\label{fig:div} Normalized waiting time $t_{\mathrm{w},\infty}/\tau_0$ (closed) after which networks no longer yield and the normalized long-time yield plateau $\Delta t_{\mathrm{y},\infty}/\tau_0$ (open) as a function of normalized stress $\sigma/\sigma_\mathrm{c}$ for networks with different critical stresses $\sigma_\mathrm{c}$, corresponding to $0.013 \leq \phi \leq 0.027$. Solid and dashed curves are predictions from eq.\ \ref{eq:PragerModel} and a self-healing active bond rupture model \cite{Mora2011}, respectively. \textit{Inset}: Normalization time $\tau_0$ as a function of $\sigma_\mathrm{c}$. Solid line is a guide to the eye.}
\end{figure}

We now focus on the mechanisms underlying the yielding and recovery processes. Based on the creep experiments and the bifurcated aging response, we expect that the fibrillar networks yield by concentrating stress into a shear band \cite{Berret2001,Chevalier2013}, similar to the behavior of networks of associative polymers \cite{Skrzeszewska2010,Leocmach2014} and microgel suspensions \cite{Meeker2004}. The boundary of the shear bands represents a local minimum in entanglement density that allows the sample to flow. If the network is allowed to rest after yielding, individual fibers dynamically rebuild the entanglement network across the boundary to restore the physical properties of the bulk sample. This picture is reminiscent of the welding of two polymer melts \cite{Prager1981,Wool1989}. The kinetics of this welding behavior is modeled by chains transporting across an interface through reptative modes so that the planar chain density $\rho$ follows
\begin{equation}
\label{eq:PragerModel}
\begin{split}
\rho(t)/\rho_\infty = & \frac{2}{\sqrt{\pi}} \left(\tau^{1/2} + 2 \sum^\infty_{k = 1} (-1)^k \times \right. \\
&\left. \left[ \tau^{1/2} \exp(-k^2/\tau) - \sqrt{\pi} k \: \mathrm{erfc}(k/\tau^{1/2})\right] \right),
\end{split}
\end{equation}
where $\tau = 2 t_\mathrm{w}/ \tau_0 N^2$, $\tau_0$ is a time scale related to the diffusion of a segment, $N$ is the number of segments per chain, and $\rho_\infty$ is the chain density at equilibrium \cite{Prager1981}. Because the mechanical modulus of the network does not significantly change across repeated yieldings, we expect that CNF disentangle under flow rather than undergo scission. Creating an interface within the CNF network via disentanglement requires individual chains to be pulled out of the network. Each chain experiences a frictional drag $\zeta$ as it is pulled out so that the energy required to create an interface can be expressed as $G_\mathrm{c} \sim \zeta \rho $  \cite{DeGennes1981,Prentice1983} Using Griffith's fracture criteria \cite{Griffith1921}, fracture energy is related to fracture stress according to $G_\mathrm{c} = \pi a \sigma_\mathrm{c}^2/E$, where $a$ is related to the initial size of the fracture and $E$ is the Young's modulus of the network. Combining these expressions leads to $\sigma(t)/\sigma_\mathrm{c} \sim (\rho(t)/\rho_\infty)^{1/2}$, which accurately fits the restructuring kinetics of the CNF networks across many orders of magnitude in time and an order of magnitude in network strength (Fig.\ \ref{fig:div}) with only one floating parameter, $\tau_0$. Importantly, this expression correctly captures the divergence of $t_\mathrm{w,\infty}$ as $\sigma$ approaches $\sigma_\mathrm{c}$. These restructuring kinetics contrast strongly with the rapid self-healing dynamics of other yield-stress fluids such as chemically crosslinked polymer networks \cite{Wei2014,Basak2014,Shi2015} or colloidal gels \cite{Wang2011,Diba2017} in which the self-healing mechanisms are controlled by the diffusion of molecular crosslinkers and chemical reaction rates. 

At higher stresses $\sigma > \sigma_\mathrm{c}$, the applied stress is large enough to always yield the sample regardless of $t_\mathrm{w}$. In this regime, yielding dynamics represent a competition between the rate of disentanglement, controlled by the applied stress, and the rate of creation of new entanglements, controlled by fibril dynamics. Because the networks are weaker immediately after yielding, we focus on the magnitude of the yield time $\Delta t_{\mathrm{y},\infty}$ measured after long $t_\mathrm{w}$ as representative of an equilibrated sample. For high stress $\sigma \gg \sigma_\mathrm{c}$, $\Delta t_{\mathrm{y},\infty}$ exhibits an exponential decay with increasing $\sigma$ (Fig.\ \ref{fig:div}). As $\sigma$ approaches $\sigma_\mathrm{c}$, $\Delta t_{\mathrm{y},\infty}$ diverges because the network can fully support the stress. There are a variety of models \cite{Zhurkov1984,Vanel2009,Skrzeszewska2010, Sprakel2011} that predict yield time as a function of applied stress for disordered materials. Many of these models assume that the network is formed by bonds that dissociate over time to allow the material to relax. This bond dissociate rate is exponentially enhanced by applied stresses so that the material fails faster at high stresses. The model is agnostic to physicochemical nature of the bond. While the exponential decay of $\Delta t_\mathrm{y,\infty}$ is observed for CNF networks at large stresses (Fig.\ \ref{fig:div}), the yield time diverges near the $\sigma_\mathrm{c}$ rather than approaching finite value. A recently proposed self-healing activated bond rupture (SH-ABR) model \cite{Mora2011,Ligoure2013}, however, accounts for this yield time divergence by incorporating the ability of the material to reform entanglements. This SH-ABR model captures the dependence of $\Delta t_{\mathrm{y},\infty}$ on $\sigma$ across orders of magnitude in time and accurately predicts the divergence near $\sigma_\mathrm{c}$ and the transition to an exponential decay when $\sigma \gg \sigma_\mathrm{c}$ (Fig.\ \ref{fig:div}). Thus, fibrillar networks yield because the rate of chain disentanglement is stress-enhanced beyond the rate at which they reform.

Extracellular matrices surrounding animal cells (e.g. collagen) similarly yield and recover, but the underlying mechanisms contrast with the cellulose model presented in this study. Our cellulose networks yield due to fibril disentanglement but can recover over long timescales with dynamic reconstruction of the entangled networks. Conversely, yielding in collagen and fibrin matrices is typically associated with plastic deformation and rupture of fibril bundles and dissociation of crosslinks \cite{Kurniawan2016}. Matrix lengthening also drives collagen fiber realignment and densification, facilitating material recovery via the formation of new crosslinks between adjacent fibers \cite{Ban2018, Ban2019}. Akin to our cellulose gels following re-entanglement, a matrix with newly formed crosslinks has a configuration that differs from its initial reference state. From these differences, the kinetics of the CNF reentanglement, described by the relaxation time $\tau_0$, should relate to the segmental dynamics of the individual fibrils between entanglement points \cite{Prager1981}. As the network stiffness $\sigma_\mathrm{c}$ increases, $\tau_0$ decreases (inset to Fig.\ \ref{fig:div}), indicating that entanglements are reformed faster as $\phi$ increases. Because entanglements require at least two fibril segments to come in contact, the rate of entanglement formation should depend on two factors -- the distance between fibrils and the fibril dynamics. As $\phi$ increases, the distance that fibril need to travel until they reach each other is decreased, increasing the collision frequency between fibrils and enhancing the rate of network formation. This behavior is qualitatively similar to that observed for peptide hydrogels \cite{Veerman2006}, gelatin networks \cite{Bohidar1993}, and colloidal gels \cite{Sandkuhler2004,Wu2013}. Due to the extended nature of cellulose fibrils, there are multiple dynamic modes controlling their motion, including center-of-mass diffusion and segmental fluctuations. Although terminal relaxations (i.e. diffusive modes) slow with increasing $\phi$, \emph{segmental} dynamics accelerate with $\phi$ because the network correlation length decreases, as observed in semidilute polymer solutions \cite{Munch1977,Rubinstein2003}. Thus, the decrease in $\tau_0$ with $\sigma_\mathrm{c}$ reflects the importance of segmental relaxations of the CNF and the decreasing distance between fibers. These mechanisms suggest that the CNF dynamics control the rate of formation of physical entanglements before and after yielding, in contrast to the mechanisms described for other structural biomacromolecules such as collagen.

In conclusion, we report the mechanical properties of cellulose nanofibril networks across the yield transition and elucidate the mechanisms of repair after yielding. Immediately after yielding, the fibrillar networks are weak and will readily yield again but after sufficient waiting time, the fibrillar network fully restores. The yielding kinetics are controlled by the competition between the breaking and formation of physical entanglements that result in a divergent yield time at the critical yield stress. This yield time exponentially decays at larger stress. We also demonstrate that the kinetics of network restructuring is consistent with a picture in which segmental dynamics of fibrils dictate recovery of entanglement density across a putative shear band. From this work we identify a new and robust method to determine the yield stress $\sigma_\mathrm{c}$ from the bifurcation in the material's aging response, and we elucidate the physical mechanisms underlying the yield transition in fibrillar networks.

\section{Acknowledgments}
R. P.-S. and C. O. acknowledge Dr. Paul Janmey for useful discussions and support from the Center for Engineering MechanoBiology, an NSF Science and Technology Center, under grant agreement CMMI: 15-48571.

\bibliography{biblio}

\end{document}